\documentclass[aps,prl,numerical,preprint,superscriptaddress,longbibliography]{revtex4-1}
\usepackage{color}
\usepackage[version=3]{mhchem} % Formula subscripts using \ce{}
\usepackage{gensymb}
\usepackage{amssymb}
\usepackage{upgreek}
\usepackage{mathtools}
\usepackage[usenames,dvipsnames]{xcolor}
\begin{document}

\title {Manipulating the Charge State of Spin Defects in Hexagonal Boron Nitride}

\author{Angus Gale}
\thanks{These authors contributed equally to this work.}
\affiliation{School of Mathematical and Physical Sciences, Faculty of Science, University of Technology Sydney, Ultimo, New South Wales 2007, Australia}

\author{Dominic Scognamiglio}
\thanks{These authors contributed equally to this work.}
\affiliation{School of Mathematical and Physical Sciences, Faculty of Science, University of Technology Sydney, Ultimo, New South Wales 2007, Australia}

\author{Ivan Zhigulin}
\affiliation{School of Mathematical and Physical Sciences, Faculty of Science, University of Technology Sydney, Ultimo, New South Wales 2007, Australia}

\author{Benjamin Whitefield}
\affiliation{School of Mathematical and Physical Sciences, Faculty of Science, University of Technology Sydney, Ultimo, New South Wales 2007, Australia}

\author{Mehran Kianinia}
\affiliation{School of Mathematical and Physical Sciences, Faculty of Science, University of Technology Sydney, Ultimo, New South Wales 2007, Australia}
\affiliation{ARC Centre of Excellence for Transformative Meta-Optical Systems (TMOS), University of Technology Sydney, Ultimo, New South Wales 2007, Australia}

\author{Igor Aharonovich}
\affiliation{School of Mathematical and Physical Sciences, Faculty of Science, University of Technology Sydney, Ultimo, New South Wales 2007, Australia}
\affiliation{ARC Centre of Excellence for Transformative Meta-Optical Systems (TMOS), University of Technology Sydney, Ultimo, New South Wales 2007, Australia}

\author{Milos Toth}
\email[]{milos.toth@uts.edu.au}
\affiliation{School of Mathematical and Physical Sciences, Faculty of Science, University of Technology Sydney, Ultimo, New South Wales 2007, Australia}
\affiliation{ARC Centre of Excellence for Transformative Meta-Optical Systems (TMOS), University of Technology Sydney, Ultimo, New South Wales 2007, Australia}

\date{May 6, 2023}
%\date{\today}

\begin{abstract}
Negatively charged boron vacancies (\ce{V_B-}) in hexagonal boron nitride (hBN) have recently gained interest as spin defects for quantum information processing and quantum sensing by a layered material. However, the boron vacancy can exist in a number of charge states in the hBN lattice, but only the -1 state has spin-dependent photoluminescence and acts as a spin-photon interface. Here, we investigate charge state switching of \ce{V_B} defects under laser and electron beam excitation. We demonstrate deterministic, reversible switching between the -1 and 0 states ($\rm V_B^-\xrightleftharpoons[]{} V_B^0 + e^-$), occurring at rates controlled by excess electrons or holes injected into hBN by a layered heterostructure device. Our work provides a means to monitor and manipulate the \ce{V_B} charge state, and to stabilize the -1 state which is a prerequisite for optical spin manipulation and readout of the defect.
\end{abstract}

%\pacs{\color{blue}61.80.-x, 79.20.Rf, 81.16.-c}

\maketitle 

Optically-active spin defects in wide bandgap materials hold promise as qubits for quantum information processing, and quantum sensing with nano-scale spatial resolution. Examples include the nitrogen-vacancy (\ce{NV-}) and silicon-vacancy (\ce{SiV^-}) centers in diamond, which have been used to achieve coherent control and manipulation of spin states \cite{Atature:2018,PRL183602,science1919,NcomAhar}. 

Hexagonal boron nitride (hBN) has recently emerged as a compelling wide bandgap van der Waals host of optically-active spin defects \cite{Gottscholl:2020,NMathBN,Ramsay2023, Stern2022, Gao20221024,Healey:2023, Huang:2022}. The most-studied spin defect in hBN is \ce{V_B-}, a negatively-charged boron monovacancy with a ground state electron spin of 1 \cite{Gottscholl:2020,Ivady:2020, Gao:2021}. The ground state electron spin can be initialised optically, manipulated by a microwave field, and read out using a spin-dependent PL emission at $\sim 800$~nm (i.e., using the optically detected magnetic resonance (ODMR) technique). However, like other spin defects in wide bandgap materials \cite{Weber:2010,NL2c04514,PRB035308}, the \ce{V_B} defect can exist in a number of charge states \cite{Weston:2018}, only one of which is suitable for spin-based applications. It is therefore important to understand \ce{V_B} charge state dynamics and stability during optical excitation, and in environments encountered in devices and sensing applications. This is, however, challenging because, in contrast to other spin defects such as the \ce{NV^{-/0}} and \ce{SiV^{-/0}} centers in diamond, only the -1 state of \ce{V_B} has a known characteristic PL emission that can be monitored experimentally.

Here, we use optical excitation and concurrent electron beam irradiation to demonstrate switching as well as reversible, deterministic manipulation of the charge states of \ce{V_B} defects in hBN. Moreover, we employ a layered heterostructure device to show that the switching is between the -1 and 0 states (i.e., $\rm V_B^-\xrightleftharpoons[]{} V_B^0 + e^-$), and provide a means to stabilize the $-1$ state which is a pre-requisite for the defect to act as a spin-photon interface. Our results provide insights into \ce{V_B} charge state dynamics during optical excitation, and in external environments such as electric fields and ionizing radiation that are encountered in devices and in real-world sensing applications. 

\begin{figure*}[b]
\resizebox{\textwidth*5/11*2}{!}{\includegraphics{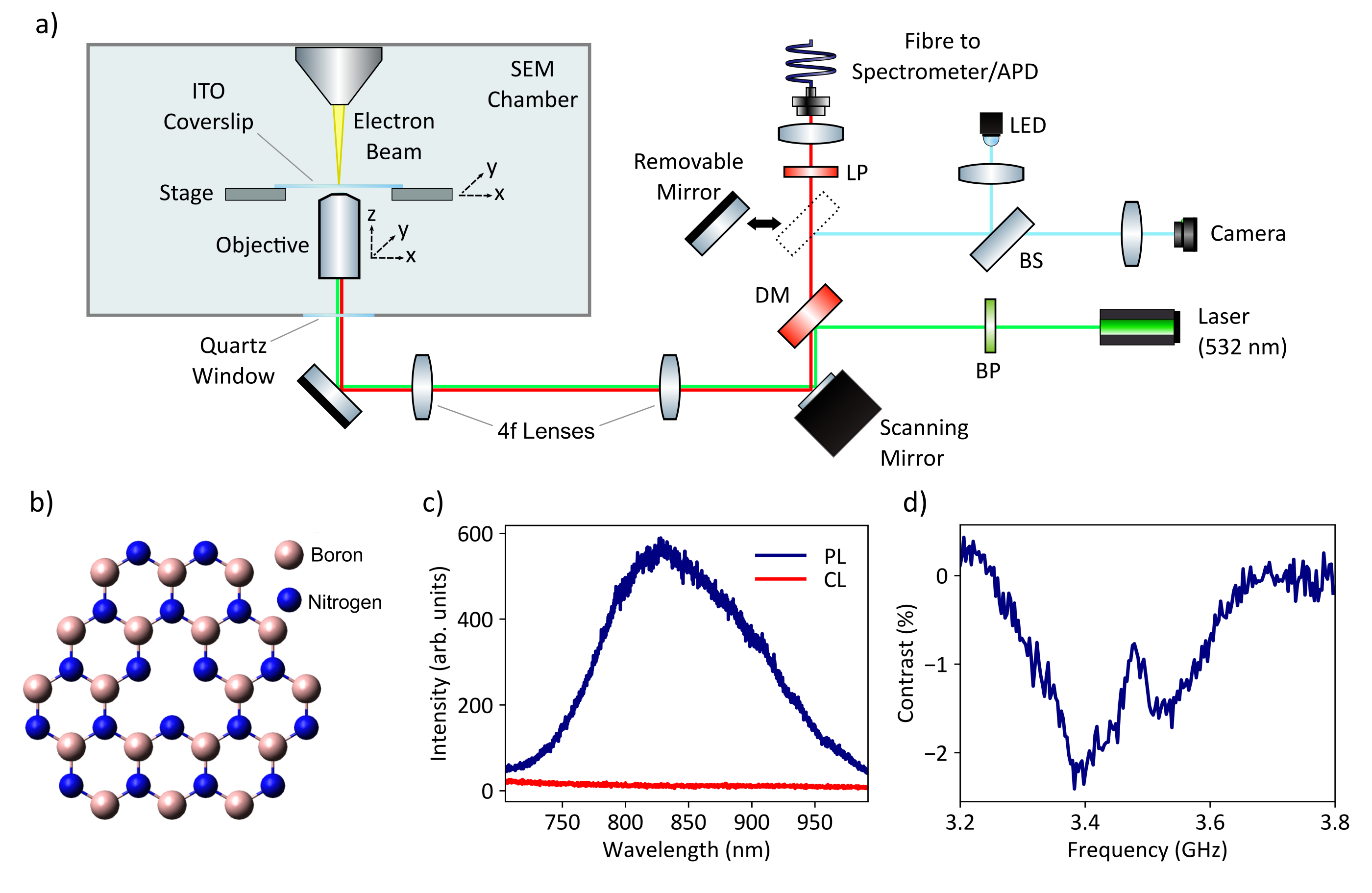}}
\caption{Overview of experimental setup and \ce{V_B} defects in hBN. a) Schematic illustration of the experimental setup. A 5~keV electron beam (yellow) irradiates hBN supported by a coverslip coated with indium tin oxide (ITO). An objective delivers a laser (green) that is coincident with the electron beam, and collects light (red) emitted by hBN. BP = Band Pass Filter, BS = Beam Splitter, DM = Dichroic Mirror, LP = Long Pass Filter, APD = Avalanche Photodiode. b) Ball and stick model of the \ce{V_B} defect in a monolayer of hBN. c) Representative PL spectrum (blue) from an ensemble of \ce{V_B-} defects excited by a 532~nm laser in the absence of the electron beam. Also shown is a CL spectrum (red) acquired from the same region in the absence of laser excitation (electron beam energy = 5~keV, electron flux = $\rm2.9 \times 10^{16}~e^-/cm^2/s$). d) Representative ODMR spectrum from an ensemble of \ce{V_B-} defects.}
\label{Fig_Overview}
\end{figure*}

A schematic of the experimental setup is shown in Fig. \ref{Fig_Overview}(a) -- a confocal laser scanning PL system incorporated in a scanning electron microscope. The setup enables coincident irradiation of a sample by an electron beam (yellow) and a laser (green), during confocal PL (red) analysis. Samples used in this work are hBN flakes that had been irradiated by 30~keV nitrogen ions to fabricate \ce{V_B} defects (details of the sample preparation and experimental methods are provided in the Supporting Information.) A schematic illustration of a \ce{V_B} defect in a single sheet of hBN is shown in Fig. \ref{Fig_Overview}(b). In the -1 charge state, \ce{V_B^-} defects are characterized by a broad PL emission at $\sim 800$~nm (blue spectrum in Fig. \ref{Fig_Overview}(c)) and a zero-field ground state splitting of $\sim 3.5$~GHz between the $\rm m_s=0$ and $\rm m_s= \pm 1$ spin states (Fig. \ref{Fig_Overview}(d))\cite{Gottscholl:2020}.

%\onecolumngrid

%\onecolumngrid
\begin{figure*}[h!b]
\resizebox{\textwidth*5/11*2}{!}{\includegraphics{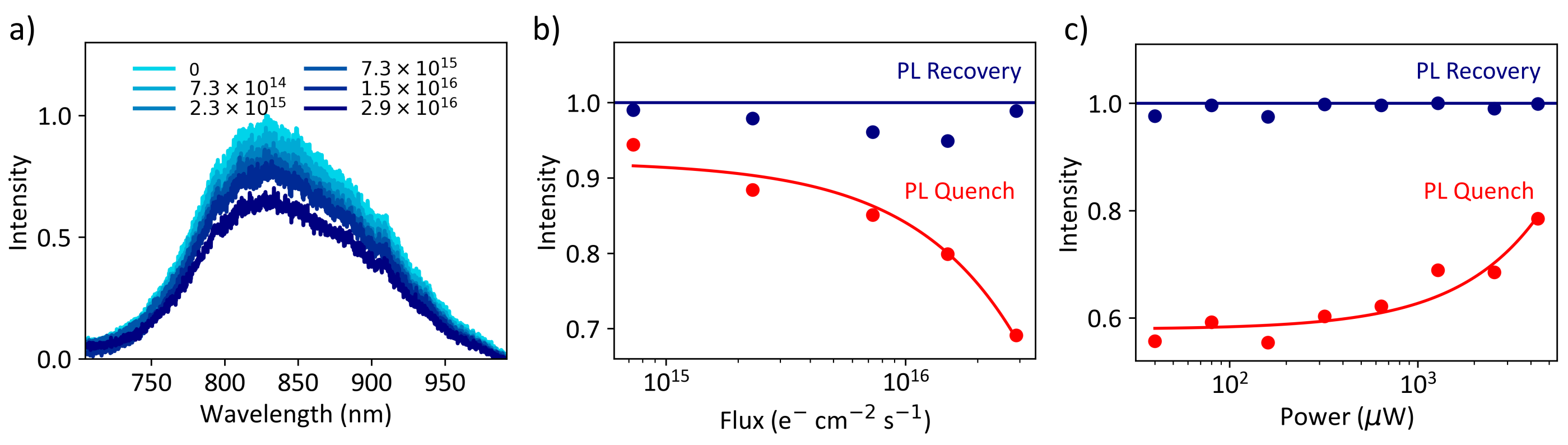}}
\caption{Charge state manipulation of \ce{V_B} defects by electron beam and laser irradiation. a) PL spectra of a \ce{V_B-} ensemble acquired as a function of electron beam flux (the fluxes used are shown in the legend in units of \ce{e^-/cm^2/s}). The spectra are normalized to the spectrum measured at zero flux (i.e., in the absence of the electron beam). They demonstrate quenching of the \ce{V_B-} PL emission by a 5~keV electron beam. b) Normalized \ce{V_B-} PL intensity plotted as a function of electron flux (red; laser power = $\rm 400~\mu W$). The PL intensity is normalized to that at zero flux. Also shown is the ratio of PL intensity measured before and after electron irradiation at each electron flux (blue), demonstrating that the PL intensity recovers when the electron beam is turned off -- i.e., the PL quenching caused by the electron beam is reversible. c) Normalized \ce{V_B-} PL intensity plotted as a function of laser power (red). In this measurement, the sample was co-irradiated by an electron beam with a flux of $\rm 7.3 \times 10^{15}~e^-/cm^2/s$. The PL intensity at each laser power is normalized to that measured at zero flux. Also shown is the ratio of PL intensity measured before and after electron irradiation at each laser power (blue), demonstrating that the PL quenching caused by the electrons is reversible. The circles and lines in b) and c) are experimental data and guides to the eye, respectively.}
\label{Fig_Curves}
\end{figure*}
%\twocolumngrid

We start by investigating the effects of a 5~keV electron beam on PL from an ensemble of \ce{V_B-} defects excited by a 532~nm laser. First, we note that electron irradiation did not generate any detectable \ce{V_B-} cathodoluminescence (CL), as is shown by the red spectrum in Fig. \ref{Fig_Overview}(c), acquired in the absence of the laser. The electrons do, however, cause quenching of the \ce{V_B-} PL signal, as is illustrated in Fig. \ref{Fig_Curves}(a) by six PL spectra acquired versus the flux (\ce{e^-/cm^2/s}) of the electron beam. The spectra are normalized to the PL spectrum measured at zero flux (i.e., in the absence of the electron beam). The \ce{V_B-} PL intensity decreases with electron flux, whilst the spectral shape remains unchanged. Moreover, the quenching is reversible, as is demonstrated in Fig. \ref{Fig_Curves}(b) by the following two plots. The first, in red, is the \ce{V_B-} PL intensity plotted versus electron flux, normalized to the PL intensity at zero flux. The plot shows that the quenching is relatively weak $(\sim 10\%)$ when the electron flux is low ($\sim 10^{15}$~\ce{e^-/cm^2/s}), and that it increases to over $30\%$ as the electron flux is increased to $\sim 3 \times 10^{16}$~\ce{e^-/cm^2/s}. The second plot in Fig. \ref{Fig_Curves}(b) is the ratio of PL intensity measured before and after irradiation by the electron beam, at each electron flux (blue). It is constant at $\sim 1$, showing that the quenching is not permanent -- i.e., the PL intensity recovers when the electron beam is turned off, at all values of electron flux used in the experiment.

Next we investigated the effect of the PL excitation laser power on the PL quenching caused by the electron beam. Fig. \ref{Fig_Curves}(c) shows a plot of the \ce{V_B-} PL intensity as a function of laser power (red). In this measurement, the sample was irradiated by a 5~keV electron beam (flux=$\rm 7.3 \times 10^{15}~e^-/cm^2/s$), and the plotted PL intensity is normalized to that measured at each laser power in the absence of the electron beam. The quenching caused by the electron beam is inhibited by the laser -- for example, at a relatively low laser power of $\rm \sim 10^2~\mu W$, the electron beam reduced the PL intensity by $\sim 40\%$, whilst at a higher laser power of $\rm \sim 4 \times 10^3~\mu W$ the same electron beam reduced the PL intensity by $\sim 20\%$. For completeness, Fig. \ref{Fig_Curves}(c) shows a plot of the ratio of PL intensity measured at each laser power before and after electron irradiation. The plot confirms that the PL quenching caused by electrons is reversible at all laser powers employed in this work.

To summarize, the electron beam causes quenching of \ce{V_B-} PL intensity by an amount that scales with electron flux (Fig. \ref{Fig_Curves}(b)) and reciprocal power of the PL excitation laser (Fig. \ref{Fig_Curves}(c)). The quenching is reversible (i.e., the PL intensity is equal before and after electron beam exposure) at all laser powers and electron fluxes investigated in this work. The electron and laser beams therefore drive two competing, rate-limited processes that quench and recover the \ce{V_B-} PL emission, respectively. The PL quenching rate scales with electron flux, the PL recovery rate with laser power, and the steady state PL intensity is determined by these two parameters. We attribute the quenching and recovery to ionization of \ce{V_B-} centers and electron-back transfer to \ce{V_B^0} defects, as is discussed below. 

\begin{figure*}[b]
\resizebox{\textwidth*5/11*2}{!}{\includegraphics{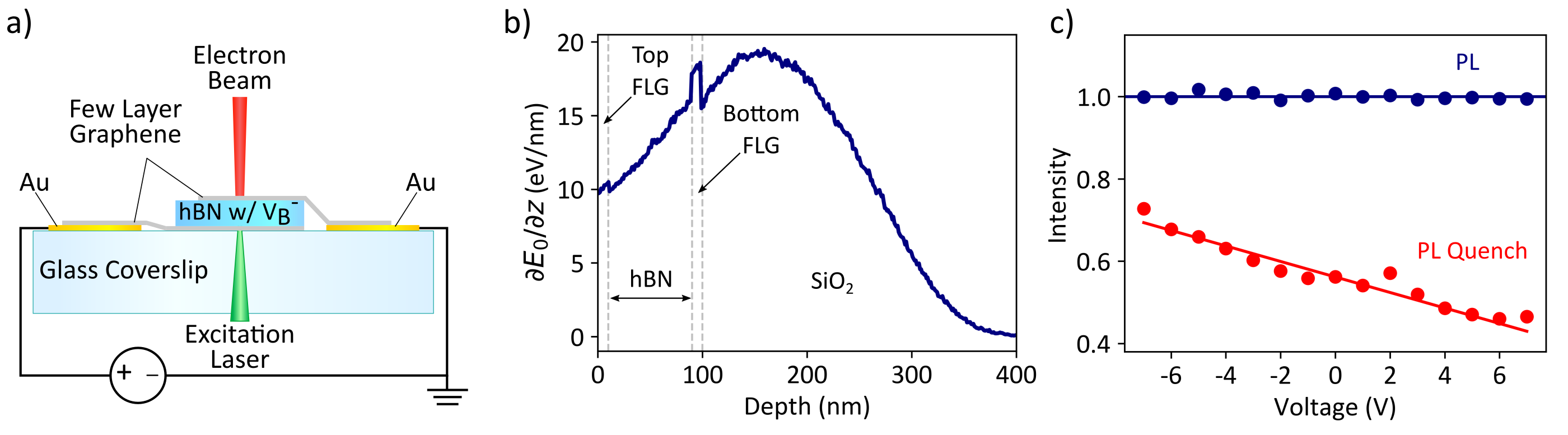}}
\caption{Electrical modulation of the \ce{V_B} charge state during co-irradiation by an electron beam and a laser. a) Cross-sectional schematic of a device comprised of an 80~nm flake of hBN encapsulated by 10~nm FLG electrodes. The device is supported by a glass coverslip substrate. The hBN is excited by a 532~nm laser from the bottom, and irradiated by a 5~keV electron beam from the top. b) Simulated depth distribution of the energy loss rate $(\partial E/\partial z)$ of a 5~keV electron incident on the device. At each depth $z$, $\partial E/\partial z$ is proportional to the generation rate of electron-hole pairs by a 5~keV electron. The range of 5~keV electrons is $\sim 350$~nm. The vertical lines show FLG interfaces. The top FLG electrode is at the surface, and the bottom electrode is at a depth of 90~nm. c) Normalized \ce{V_B-} PL emission intensity plotted versus bias (red). The PL intensity was measured during irradiation by electrons. The device bias modulates the magnitude of PL quenching caused by the electron beam. Also shown is the PL intensity measured at each bias in the absence of the electron beam (blue), showing that the device bias has no effect on PL intensity in the absence of electron irradiation. The circles and lines are experimental data and guides to the eye, respectively.}
\label{Fig_Device}
\end{figure*}

Next, we used the layered device shown schematically in Fig. \ref{Fig_Device}(a) to inject carriers into hBN and investigate their effects on the PL quenching effect. The device consists of gold contacts and a pair of few-layer graphene (FLG) electrodes that encapsulate an 80~nm flake of hBN (an optical image and PL spectra of the device are shown in Fig. S1 of the Supporting Information). The top FLG electrode was grounded, and a bias voltage was applied to the bottom electrode. The device was loaded into the SEM and irradiated from the top by a 5~keV electron beam and from the bottom by a 532~nm PL excitation laser. The 5~keV electrons penetrate through the FLG/hBN/FLG device, into the substrate, as is shown in Fig. \ref{Fig_Device}(b) by a plot of $\frac{\partial E}{\partial z}$ versus depth $(z)$. $\frac{\partial E}{\partial z}$ was simulated using the Monte Carlo package CASINO \cite{Drouin:2007}. It is the mean energy lost by each electron per unit distance travelled through the sample. The area under the curve $(E)$ is the total energy deposited into the sample per electron, and it falls off to zero at a depth of $\sim 350$~nm, the maximum penetration range of the electron beam.

%\onecolumngrid
%\twocolumngrid

The device bias modulates the magnitude of the \ce{V_B-} PL quenching caused by the  electron beam. Specifically, as is shown by the red curve in Fig. \ref{Fig_Device}(c), application of a positive or a negative bias to the bottom FLG electrode enhances or inhibits the quenching, respectively. In this measurement, the laser power and electron beam flux were both fixed, and we varied only the bias applied to the bottom FLG electrode -- for example, at a bias of -6~V, the electron beam reduced the \ce{V_B-} PL intensity by $\sim 30\%$, whilst at +6~V, it reduced the PL intensity by $\sim 57\%$. Conversely, in the absence of electron irradiation, the device bias has no effect on the \ce{V_B-} PL intensity. This is demonstrated by the blue plot in Fig. \ref{Fig_Device}(c), which shows that the PL intensity does not change with bias, when measured in the absence of the electron beam,.

We now turn to the mechanisms behind the \ce{V_B-} emission quenching effect. Three potential explanations are that the electron beam: (i) restructures the atomic configuration of \ce{V_B} defects or generates new lattice defects through “electron-beam-damage” mechanisms analogous to those reported in other dielectrics \cite{Cazaux:1999}, (ii) promotes relaxation of excited \ce{V_B-} centers through a nonradiative pathway such as the intersystem crossing \cite{Gottscholl:2020} responsible for \ce{V_B-} ODMR contrast, or (iii) changes the charge state of \ce{V_B-} centers. Of these, our results are consistent only with the latter – specifically, the PL quenching is due to ionization (i.e., $\rm V_B^- \longrightarrow V_B^0$) caused by the electron beam, and the PL recovery is due to a photo-activated electron back-transfer process (i.e., $\rm V_B^0\longrightarrow V_B^-$) driven by the 532~nm laser. To substantiate this claim, we consider in detail the operation of the device used in Fig. \ref{Fig_Device}. 

\begin{figure*}[b]
\resizebox{\textwidth*5/11*20/9}{!}{\includegraphics{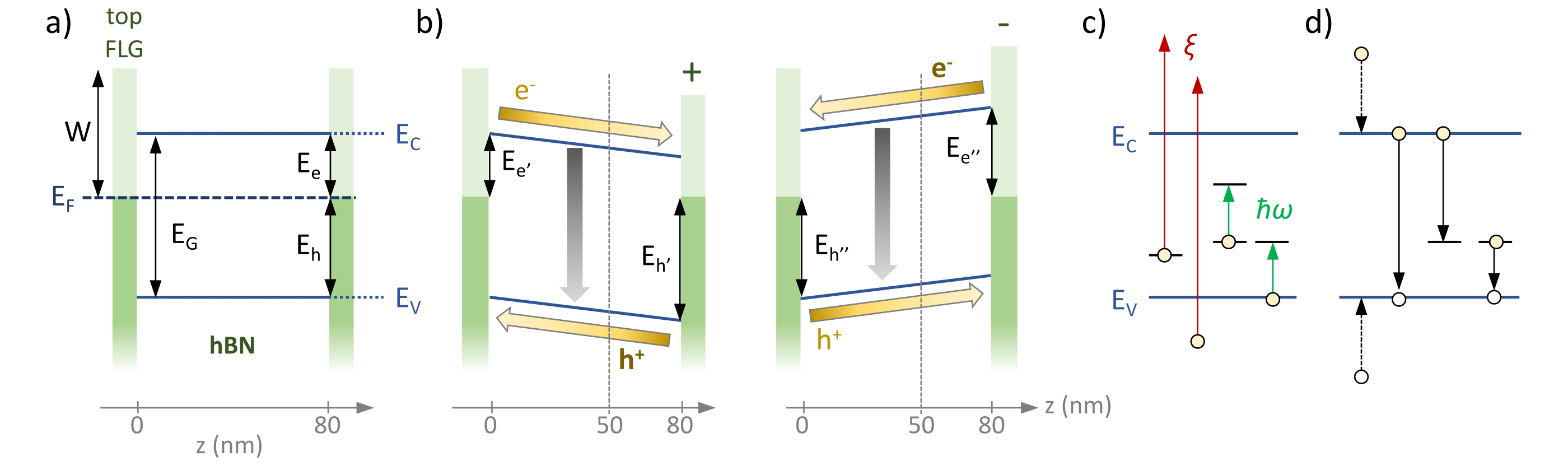}}
\caption{Simplified electron energy diagrams. a) The device in Fig. \ref{Fig_Device} with both electrodes grounded. b) The device with positive (left) and negative (right) bias applied to the bottom FLG electrode. The yellow arrows indicate the flow directions of carriers in hBN. The gray arrows represent electron-hole recombination occurring throughout the hBN layer (Eqn. \ref{Eqn_rec}), and z is distance below the top FLG electrode. The concentration of \ce{V_B} defects has a minimum at $\rm z \approx 0$ and a maximum at $\rm z \approx 50~nm$. c) Electron-induced ionization (red) of a defect, ionization of an electron in the valence band (red), and photo-excitation (green) of a defect and an electron in the valence band. d) Examples of transitions that involve free carriers in hBN: thermalization of hot carriers to band edges, recombination of carriers across the bandgap, capture of a thermalized electron at a trap state, recombination of a trapped electron with a hole in the valence band. In  c) and d) yellow and white circles represent electrons and holes, respectively. 
}
\label{Fig_EDiagram}
\end{figure*}

Fig. \ref{Fig_EDiagram}(a) is a simplified flat band electron energy diagram of the FLG/hBN/FLG heterostructure under zero bias. \ce{E_F} is the Fermi level; W is the work function of FLG \cite{Kawano:2022}; \ce{E_G}, \ce{E_C} and \ce{E_V} are the hBN bandgap \cite{Cassabois:2016}, conduction band minimum and valence band maximum; \ce{E_e} and \ce{E_h} are the energy difference between \ce{E_F} and the hBN conduction and valence band, respectively; and $z$ is distance below the top surface of the hBN layer (i.e., `depth' in Fig. \ref{Fig_Device}(b)). Application of a positive or a negative bias voltage to the bottom FLG electrode generates applied electric fields, as is illustrated in Fig. \ref{Fig_EDiagram}(b).

%\onecolumngrid
%\twocolumngrid

We first note that the bias (of up to ±8 V) had no effect on the \ce{V_B-} PL emission in the absence of electron irradiation. The device bias does, however, modulate the \ce{V_B-} emission intensity during electron irradiation (Fig. \ref{Fig_Device}(c)). The 5 keV electrons act primarily as ionizing radiation. They penetrate the entire FLG/hBN/FLG stack (Fig. \ref{Fig_Device}(b)) and excite carriers throughout the heterostructure:
\begin{equation}
\frac{\partial n_{e,h}}{\partial t} = \frac{1}{E_p}\frac{\partial E}{\partial z}f.
\label{Eqn_ion}
\end{equation}
Here, $\frac{\partial n_{e,h}}{\partial t}$ \cite{Everhart:1971} is the generation rate per unit volume of hot carriers (both electrons $e$ and holes $h$) at depth $z$ in the heterostructure, $E_p$ is the mean energy that must be lost by the electron beam to excite one electron-hole pair \cite{Alig:1975}, $\frac{\partial E}{\partial z}$ is the mean energy lost at coordinate $z$ by a beam electron per unit distance travelled through the heterostructure (plotted for our device in Fig. \ref{Fig_Device}(b)), and $f$ is the electron beam flux.

Throughout the hBN layer, hot electrons and holes $(n_{e,h})$ are excited into the conduction and valence band, respectively, at a rate that is proportional to $\frac{\partial E}{\partial z}$. These carriers cascade and thermalize to the band edges \cite{Fitting:2011} and drift towards the FLG electrodes, as shown by the yellow arrows in Fig. \ref{Fig_EDiagram}(b) for the case of positive (left) and negative (right) bias applied to the bottom FLG electrode. Ultimately, the carriers either flow to the FLG electrodes, or recombine in the hBN layer. Recombination \cite{Fitting:2011} can be expressed as:
\begin{equation}
\frac{\partial n}{\partial t} = vn_e\sigma n_h,
\label{Eqn_rec}
\end{equation}
where $\frac{\partial n}{\partial t}$ is the carrier rate of change of concentration due to recombination (indicated by gray vertical arrows in Fig. \ref{Fig_EDiagram} (b)) at coordinate $z$ in the hBN layer, $v$ is the minority carrier thermal velocity, and $\sigma$ is the recombination cross-section. 

The carriers excited by the electron beam are excited not only in the hBN layer, but also in the FLG electrodes (as per Eqn. \ref{Eqn_ion}, and the curve of $\frac{\partial E}{\partial z}$ versus $z$ which is plotted in Fig. \ref{Fig_Device}(b)). These carriers either cascade, thermalize and recombine in FLG, or are injected into hBN over the barriers \ce{E_{e'}}, \ce{E_{h'}}, \ce{E_{e''}} and \ce{E_{h''}} shown in Fig. \ref{Fig_EDiagram}(b). In hBN, the injected carriers thermalize to the band edges and contribute to the drift and recombination currents, as is indicated by the yellow and gray arrows in Fig. \ref{Fig_EDiagram}(b). However, critically, these electrons and holes are injected from opposite sides of the hBN layer (at $z = 0$ and 80~nm), and recombine as they flow through the hBN. Consequently, the carrier concentrations vary across the hBN -- when the bottom FLG electrode is biased positive (negative), there is a net excess of holes (electrons) at the bottom (top) of the hBN layer. This asymmetry is important because the depth distribution of \ce{V_B} defects is not uniform, and a net excess of either electrons or holes is therefore supplied to the defects under different bias configurations, as is detailed below.

The depth distribution of \ce{V_B} defect in hBN is determined by the fabrication method. Here, the defects were generated by 30~keV nitrogen ions.  Their concentration has a minimum at the top surface (i.e., at $z=0$ in Fig. \ref{Fig_EDiagram}) and a maximum at $z\sim 50$~nm (see Fig. S2 of the Supporting Information) -- that is, the \ce{V_B} defects are concentrated preferentially near the bottom FLG electrode. Now, when a positive bias is applied to the bottom electrode, holes are injected into hBN from the bottom electrode, and electrons from the top electrode, as is indicated by the yellow arrows in Fig. \ref{Fig_EDiagram}(b). These carriers recombine as they flow through the hBN layer, yielding a net excess of holes in the \ce{V_B}-rich region near the bottom FLG electrode. Conversely, a negative bias gives rise to the opposite scenario -- electrons are injected into the hBN layer from the bottom electrode and holes from the top electrode. Hence, a net of excess holes (electrons) is supplied to \ce{V_B} defects when a positive (negative) bias is applied to the bottom FLG electrode, which correlates with the enhancement (inhibition) of PL quenching caused by the electron beam seen in Fig. \ref{Fig_Device}(c).

Furthermore, the excess of one carrier type (electrons or holes) near the bottom electrode is exacerbated by the fact that more carriers are exited by 5~keV electrons in the bottom electrode ($80 < z < 90$~nm) than in the top electrode ($-10 < z < 0$~nm), because $\frac{\partial E}{\partial z}$ is greater in the bottom electrode, as is seen in Fig. \ref{Fig_Device} (b). Specifically, $\frac{\partial E}{\partial z} \sim 10$~eV/nm and $\sim 18$~eV/nm in the top and bottom FLG electrodes, respectively. 

Based on the above, application of a positive (negative) bias to the bottom electrode results in a net supply of excess holes (electrons) to \ce{V_B} defects. This correlates with enhancement of the \ce{V_B^-} emission quenching (recovery) driven by the electron (laser) beam. In the presence of excess holes, the quenching is enhanced, consistent with it being caused by ionization ($\rm V_B^- \longrightarrow V_B^0 + e^-$). Conversely, in the presence of excess electrons, the quenching is inhibited, consistent with it being caused by electron back-transfer ($\rm V_B^0 + e^- \longrightarrow V_B^-$).

A consequence and a prediction of the above model is that the modulation caused by the device bias should invert if the bias configuration is reversed. That is, application of a positive (negative) bias to the top FLG electrode should inhibit (enhance) the quenching, respectively. This is the opposite of the behavior seen in Fig. \ref{Fig_Device}(c), and is indeed what is observed experimentally, as is shown in Fig. S3 of the Supporting Information.

Our results do not reveal the exact transitions responsible for the charge state switching. We can, nonetheless, discuss the various possibilities and speculate on the most likely pathways. Thus, with reference to Fig. \ref{Fig_EDiagram}(c) and (d), we have:
\begin{equation}
\rm V_B^- \xrightleftharpoons[\hbar\omega]{\xi} V_B^0 + e^-.
\label{Eqn_ionize2}
\end{equation}
The energy needed to ionize \ce{V_B-} centers $\rm (\xi)$ is provided by 5~keV electrons which can drive all allowed transitions between occupied and unoccupied states in hBN. That is, 5~keV electrons can ionize \ce{V_B-} centers as well as all electron traps and electrons in the valence band of hBN. Excitations that promote electrons into the conduction band dominate (over those that promote electrons into trap states in the bandgap) since the transition probability scales with the density of unoccupied (final) states \cite{Stobbe:1991}. Hence, the most likely \ce{V_B-} ionization pathways promote electrons into the conduction band. The electrons thus liberated via the forward reaction in Eqn. \ref{Eqn_ionize2} then cascade and thermalize to the conduction band minimum and are most likely trapped at defect states in the bandgap. 

The 532~nm laser $(\hbar\omega)$ can excite transitions between electronic states separated by up to 2.33~eV. Two examples are shown in green in Fig. \ref{Fig_EDiagram}(c). Such transitions are likely responsible for the photo-activation that leads to the formation of \ce{V_B^-} via the reverse reaction in Eqn. \ref{Eqn_ionize2}. 

Carriers that are excited by the electron beam in the FLG electrodes and injected into hBN cascade and thermalize to the band edges and recombine through pathways such as those shown in Fig. \ref{Fig_EDiagram}(d). Excess electrons promote \ce{V_B^-} formation, likely through two mechanisms. The first is capture of thermalized electrons in the conduction band by \ce{V_B^0} defects. The second is enhancement of the the laser-driven \ce{V_B-} recovery process, by populating electron traps in the vicinity of \ce{V_B^0} defects (i.e., by supplying trapped electrons for the laser-driven photo-excitation process in Eqn. \ref{Eqn_ionize2}). Similarly, excess holes likely promote \ce{V_B^0} formation, via capture of holes in the valence band by \ce{V_B^-} centers, and via depopulation of electron traps in the vicinity of \ce{V_B} defects.

Finally, we return to the observation that the device bias (of up to $\pm 8$ V) had no effect on the \ce{V_B-} emission in the absence of electron irradiation (blue curve in Fig. \ref{Fig_Device}(c)) -- that is, injection of photo-excited carriers from FLG into hBN appears to not play a role in the observed \ce{V_B-} PL emission modulation. This is not surprising since the PL excitation laser promotes \ce{V_B^-} formation, and whilst electron injection from FLG to hBN is possible, the barriers for hole injection ($\rm E_{h'}$ and $\rm E_{h''}$, shown in Fig. \ref{Fig_EDiagram}(b)) are greater than the laser energy $(\hbar\omega)$, under all bias configurations used in this work. We also return to the observation that the electron beam did not generate any detectable \ce{V_B-} CL (red spectrum in Fig. \ref{Fig_Overview}(c)) in the absence laser excitation. This is also not surprising since the electrons ionize \ce{V_B-} centers, and promote the formation of \ce{V_B^0} defects.

In conclusion, our results demonstrate reversible manipulation of the charge state of \ce{V_B} defects in hBN. \ce{V_B^-} centres are ionized by 5~keV electrons, and the charge state is recovered by photoexcitation at 532~nm. The two process rates can be modulated by injecting excess electrons or holes into hBN using a layered heterostructure device which can be used to stabilize the -1 state of the defects. 

We acknowledge financial support from the Australian Research Council (CE200100010, FT220100053) and the Office of Naval Research Global (N62909-22-1-2028).

\bibliography{VB_Charge_Refs}

\end{document}